\documentclass[12pt,preprint]{aastex}

\shorttitle{Imaging with a MCDA}

\shortauthors{Marois et al.}

\begin{document}

\title{Differential Imaging with a Multicolor Detector Assembly:\\ A New ExoPlanet Finder Concept}

\author{CHRISTIAN MAROIS\altaffilmark{1}, REN\'{E} RACINE\altaffilmark{1}, REN\'{E} DOYON\altaffilmark{1}, DAVID LAFRENI\`{E}RE\altaffilmark{1}, and DANIEL NADEAU\altaffilmark{1}}
\affil{Observatoire du Mont M\'{e}gantic and D\'{e}partement de Physique, Universit\'{e} de Montr\'{e}al, Montr\'{e}al, QC H3C 3J7, Canada}
\email{marois@astro.umontreal.ca, racine@astro.umontreal.ca, doyon@astro.umontreal.ca, david@astro.umontreal.ca, nadeau@astro.umontreal.ca}

\altaffiltext{1}{Visiting Astronomer, Canada-France-Hawaii Telescope, operated by the National Research Council of Canada, the Centre National de la Recherche Scientifique, France, and the University of Hawaii.}

\begin{abstract} 
Simultaneous spectral differential imaging is a high contrast technique by which subtraction of simultaneous images reduces noise from atmospheric speckles and optical aberrations. Small non-common wave front errors between channels can seriously degrade its performance. We present a new concept, a multicolor detector assembly (MCDA), which can eliminate this problem. The device consists of an infrared detector and a microlens array onto the flat side of which a checkerboard pattern of narrow-band micro-filters is deposited, each micro-filter coinciding with a microlens. Practical considerations for successful implementation of the technique are mentioned. Numerical simulations predict a noise attenuation of 10$^{-3}$ at 0.5$^{\prime \prime}$~for a 10$^5$ seconds integration on a $m_H$~=~5 star of Strehl ratio 0.9 taken with an 8-m telescope. This reaches a contrast of 10$^{-7}$ at an angular distance of 0.5$^{\prime \prime}$~from the center of the star image. 
\end{abstract} 


\section{Introduction and Motivation}
Direct imaging and spectroscopy of exoplanets are crucial to constrain their masses, study their atmospheres and, ultimately, find telltale signatures of life. The task is dauntingly difficult. The planet image is hugely fainter than the background from the brilliant stellar image. The most obvious problem is photon noise in the bright star point spread function (PSF). A remedy is coronagraphy \citep{lyot1933,roddier1997,rouan2000,baudoz2000,abe2001,kasdin2003}, which attenuates the coherent diffraction pattern of the on-axis PSF. At high Strehl ratios the diffraction pattern dominates much of the PSF and coronagraphs can achieve strong attenuation \citep{kilston2002}. With the relatively modest Strehl ratios achieved with current adaptive optics (AO) systems on large ground based telescopes, the efficiency of coronagraphy is limited by the presence of a rapidly changing halo of speckles generated by atmospheric phase distortion. This increases the local PSF variance by orders of magnitude above that caused by photon noise \citep{racine1999}. In addition, aberrations in the optical train, unfiltered by the AO system, produce speckles of much longer lifetime, hence much brighter in a long-exposure image, than atmospheric speckles. This noise can be dominant for integrations longer than a few seconds \citep{marois2003a,marois2004}. Quasi-static speckles also limit the high-contrast performance of the Hubble Space Telescope \citep{schneider2003}, in the form of the so-called ``breathing'' problem. All this underscores the importance of PSF calibration for high contrast imaging, which is the topic of this paper.

Simultaneous spectral differential imaging \citep{smith1987,racine1999,marois2000a,sparks2002,biller2004} is such a calibration technique. Images are acquired simultaneously at adjacent wavelengths in a spectral range where the planetary and stellar spectra differ appreciably. Judicious image subtraction removes the stellar image and conserves that of any companion. A useful spectral feature for this purpose is the sharp methane absorption band head at 1.6~$\mu$m \citep{rosenthal1996} found in relatively cold atmospheres \citep{burgasser2002} such as those of cold T type brown dwarfs and Jovian planets.

\begin{figure}
\epsscale{1}
\plotone{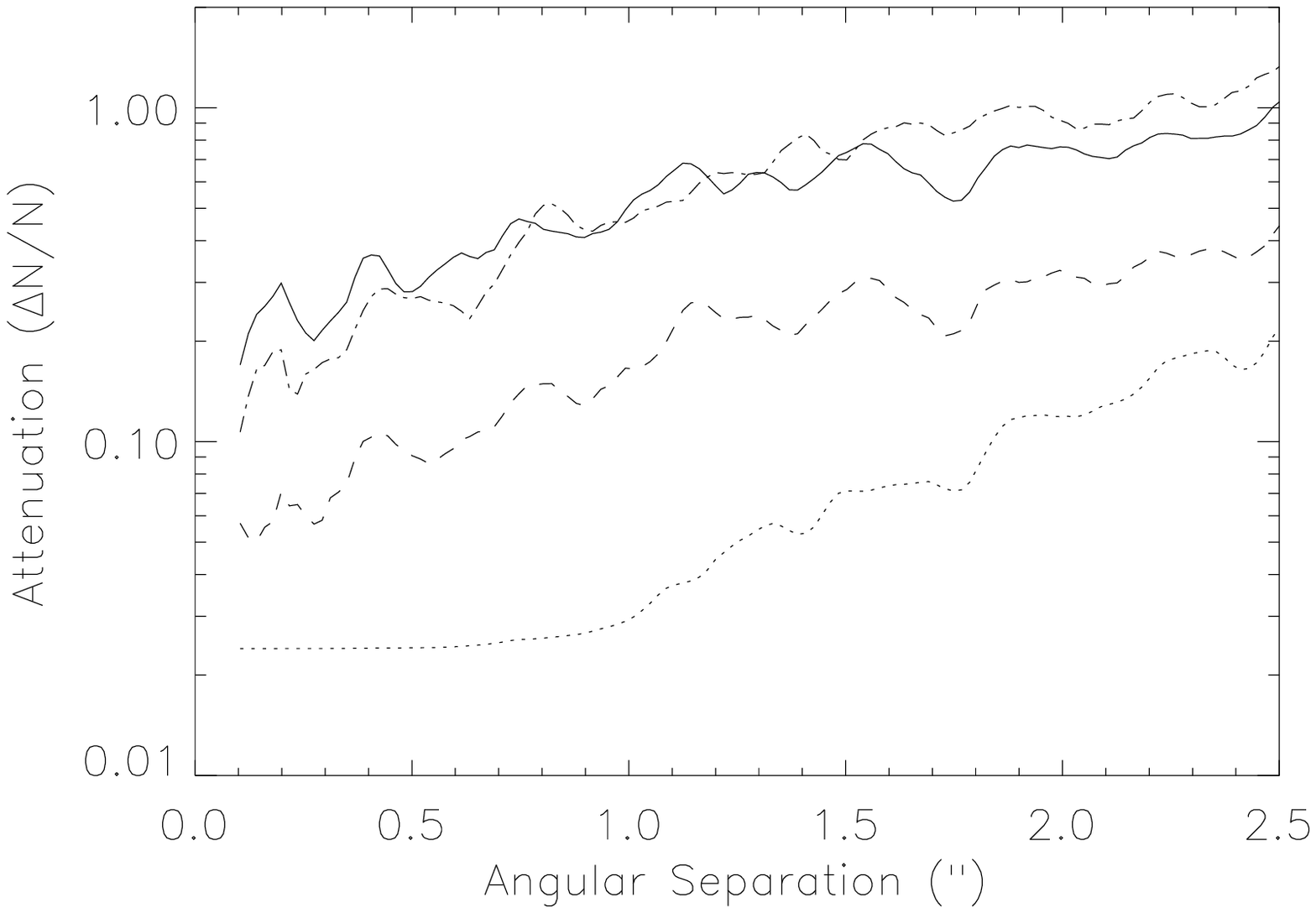}
\caption{Median of the pixel by pixel ratio of residual noise to initial noise as a function of the distance from the PSF center for (a) the difference of two simultaneous TRIDENT images (solid line), (b) the difference between differences of simultaneous TRIDENT images of two stars observed sequentially (dashed line), (c) a difference image whose noise would only consist of the speckle, read, photon and flat field noises of the TRIDENT images (dotted line) and (d) a simulated difference image with the same noises plus an RMS wave front difference between channels of $\lambda$/20 (dot-dash line). \label{fig1}}
\end{figure}

The TRIDENT camera \citep{marois2000b,marois2003a,marois2003b,marois2003c,marois2004}, a 3-channel (1.580, 1.625 and 1.680~$\mu$m, 1\% bandwidth) differential imager, has been extensively used with the AO system of the 3.6-m Canada-France-Hawaii Telescope \citep{rigaut1998}. From the data, one can determine the level of noise attenuation achieved. The noise attenuation factor $\Delta$N/N is defined as the median over an annulus of the pixel-by-pixel ratio of the absolute value of the intensity in the difference image to its absolute value in the original image after subtraction of an azimuthally averaged profile. The noise structure of the PSF at one wavelength is dominated by static aberrations and is well approximated with $\sim$~130~nm of static phase screen following a power-law distribution P~$\propto \nu^{\alpha}$ with $\alpha$~=~$-2.7$. This level of aberration is consistent with that measured (110~nm RMS) for the static component of the PUEO AO system \citep{rigaut1998}. Fig.~\ref{fig1} illustrates typical noise attenuations obtained with TRIDENT. At subarcsecond separations the noise residuals in a difference image are $\sim 10$ times larger than the sum of the photon and read noises. Results from numerical simulations indicate that $\sim$~$\lambda$/20 of non-common path aberrations can explain TRIDENT limited performance. A factor of $\sim 4$ improvement is obtained by subtraction of the residuals of a reference star observed shortly after and under the same conditions as the science target. This indicates that differential aberrations between channels contribute a large fraction of the residual noise in a difference of simultaneous images. It also appears that some of the noise after subtraction of the reference star image is caused by slight changes in optical aberrations between the target and the reference star observations.

In this paper, in light of the non-common path limitations, a new differential imager concept is introduced that should improve detection limits by a large factor. Simulations are made to estimate the performances and the expected detection limits with 8-m class telescopes and high order AO systems. 

\section{The Multicolor Detector Assembly Concept}
\begin{figure}
\epsscale{1}
\plotone{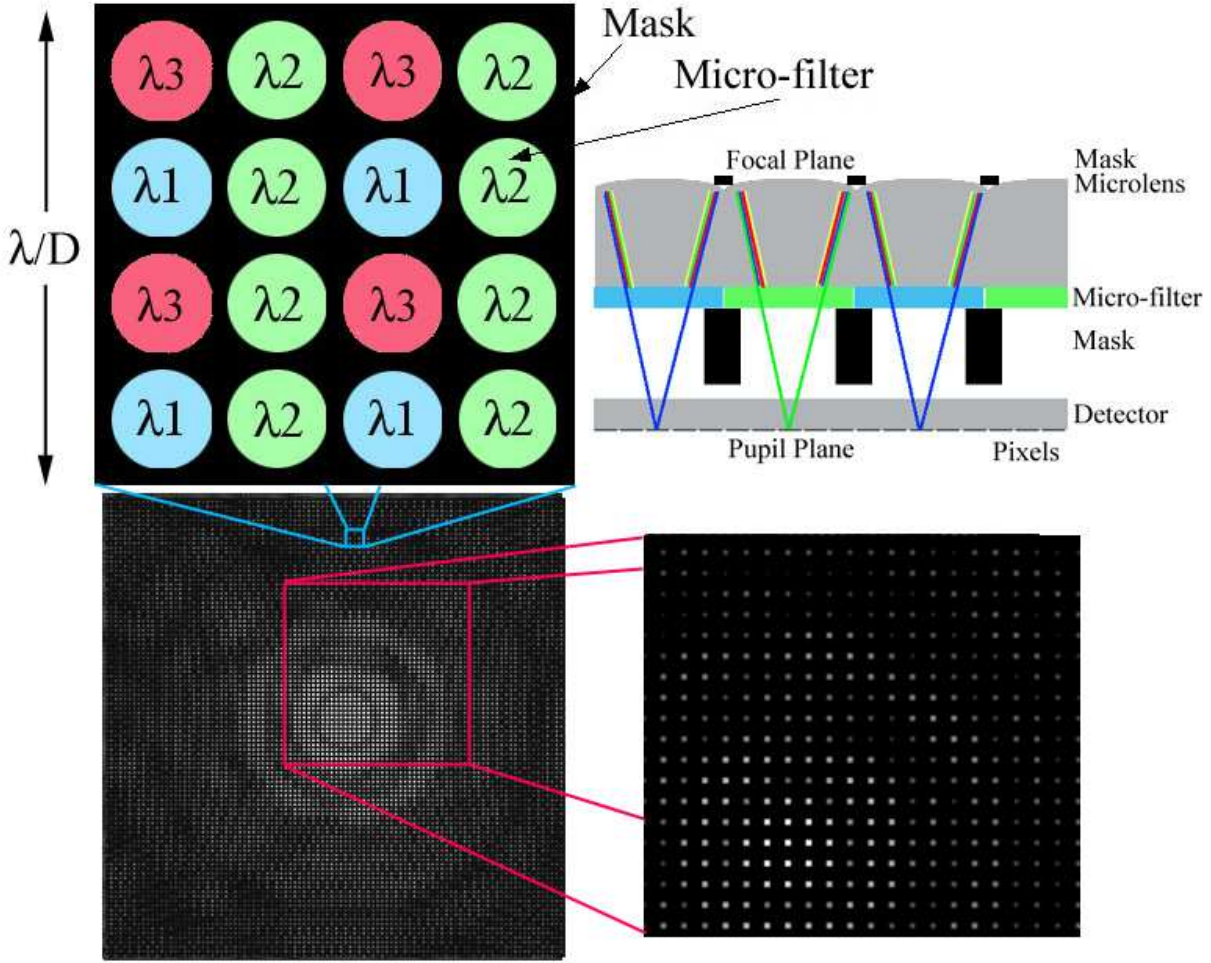}
\caption{Concept of a three wavelength differential imager with an MCDA showing the whole assembly (upper right) and the micro-filter array (upper left). Here, each microlens covers a $5 \times 5$ pixel cluster. A PSF and one of its checker board monochromatic samples are shown at the bottom. See text for details.\label{fig2}}
\end{figure}

A solution to the non-common path problem is to perform the spectral separation in the image plane of a single optical path imager with a Multicolor Detector Assembly (MCDA). A three-wavelength MCDA is illustrated in Fig.~\ref{fig2}. A microlens array located at the focal plane is used to concentrate the light of each PSF spatial sample into well separated detector pixels (each microlens covers a $5 \times 5$ pixel cluster) in order to minimize interpixel crosstalk, as discussed below. A mask on both faces of the microlens array blocks the light which may scatter between microlenses. Each microlens in a 2~$\times$~2 microlens cluster is covered with one of three different micro-filters, two microlenses having the same filter for signal-to-noise optimization. The fact that only a small central area of each micro filter is used relaxes the requirement for edge-to-edge filter uniformity. A detector read-out yields four critically sampled ``checkerboard'' images of the PSF, one at each wavelength, that can be interpolated to construct a filled image.

The attenuation of noise structures allowed by this device, in the absence of photon, read and flat field noises, can theoretically be arbitrarily large if the noise structures - speckles - are critically sampled at each wavelength over a PSF image whose extent is unlimited. This is so because, by the sampling theorem for bandwidth limited images, it is then possible to perfectly interpolate the pixel intensity at a given wavelength for any pixel which is ``blind'' at that wavelength. Complete attenuation is of course not achievable in practice because the idealized conditions posed above cannot be realized. Flat field errors will probably be the dominant limitation at high flux levels. The MCDA itself may entail limiting factors such as the two mentioned below. 

\begin{figure}
\epsscale{1}
\plotone{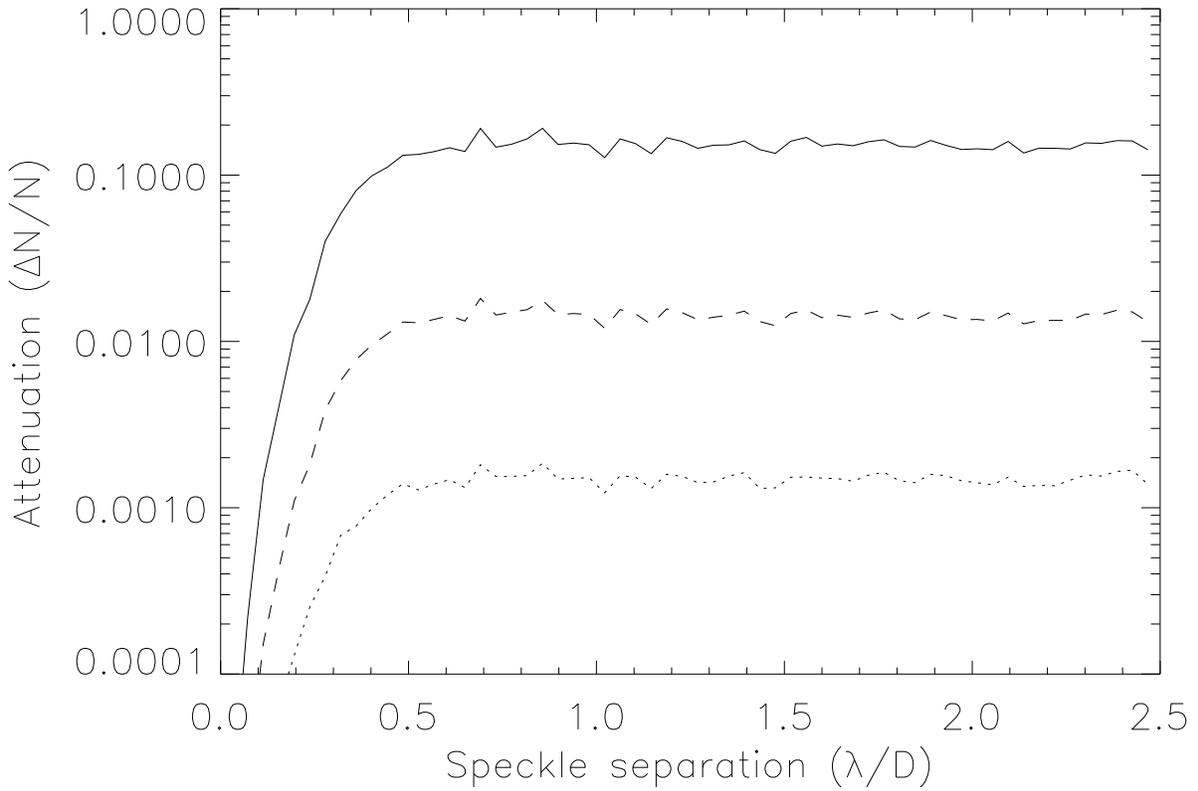}
\caption{Effect of sample to sample crosstalk in an MCDA from numerical simulations. This shows the noise attenuation achieved by the subtraction of two PSF samples as a function of speckle separation with wavelength, expressed in $\lambda$/D units, when attenuation is only limited by 0.001 (dotted line), 0.01 (dashed line) and 0.1 (solid line) of the sample to sample crosstalk.\label{fig3}}
\end{figure}

The microlens array is required to minimize signal crosstalk between PSF sampling points. With current infrared detectors, a pixel can leak as much as 40\% of its signal to its adjacent pixels \citep{finger1998}. Diffraction by the microlenses can also be a significant source of crosstalk. Crosstalk limits the PSF noise attenuation as follows. The distance of a speckle to the PSF center is proportional to wavelength. PSFs obtained at different wavelengths must be re-scaled before being subtracted. The leak of a speckle signal sampled at one wavelength to a sampling point for another wavelength no longer coincides spatially with itself after re-scaling. Subtraction of the re-scaled images leaves a residual whose intensity is equal to the crosstalk coefficient X times the difference in the speckle intensity between the radially shifted samples. The average residual noise after image subtraction then increases from zero at the PSF center, where the wavelength shift is nil, to X times its initial level when the radial shift between wavelengths exceeds the FWHM ($\sim$~$\lambda$/D) of the speckles, i.e. at an angular offset $\theta > \lambda^2$/($\Delta \lambda \times \rm{D}$). At larger offsets, the crosstalk induced noise saturates at an amplitude of X times the speckle intensity and the limiting attenuation factor becomes equal to the crosstalk level X. Numerical simulations (Fig.~\ref{fig3}) confirm this analysis. It is expected that sample separations $\sim 5 \times 18.5$~$\mu $m pixels (to reduce charge leaks between samples to $< 0.001$) and microlens f-ratios $<17$ (to limit the spillover of a pupil diffraction pattern onto an adjacent pupil image) will allow X~$< 0.001$ at $\lambda$ $\sim$~1.6~$\mu $m.\footnote{For a circular aperture, the mean intensity diffracted at radius $r$ is $\sim$~$0.04\left[ r/\left( \lambda \times f/\rm{D}\right) \right]^{-3}$ for unit central intensity.  For diffraction crosstalk to be $< 0.001$ at a wavelength of 1.6~$\mu $m and at a distance of 5$\times$18.5~$\mu $m requires $f/\rm{D}$~$<$~17.} To minimize the sample to sample crosstalk and to block gaps between microlenses, two masks are introduced above and under the microlens array.

PSF decorrelation between wavelengths can also occur with an MCDA if the optical path is chromatic. Refractive optics produce wave fronts of different dimensions at different wavelengths because of the wavelength dependence of the index of refraction. Atmospheric dispersion also decorrelates the PSFs noise structure because the wave fronts for different wavelengths follow different optical paths through the atmosphere and instrument. This leads to non-common path errors since the footprints on optical surfaces of wave fronts at different wavelengths will be sheared. Simulations indicate that a $\sim $1/1000 wave front shear limits attenuation to $\sim $10$^{-3}$. The effects of chromatism and atmospheric refraction can be mitigated by using reflective optics and atmospheric dispersion compensators.

\section{Simulated Performance with an MCDA}

\begin{figure}
\epsscale{1}
\plotone{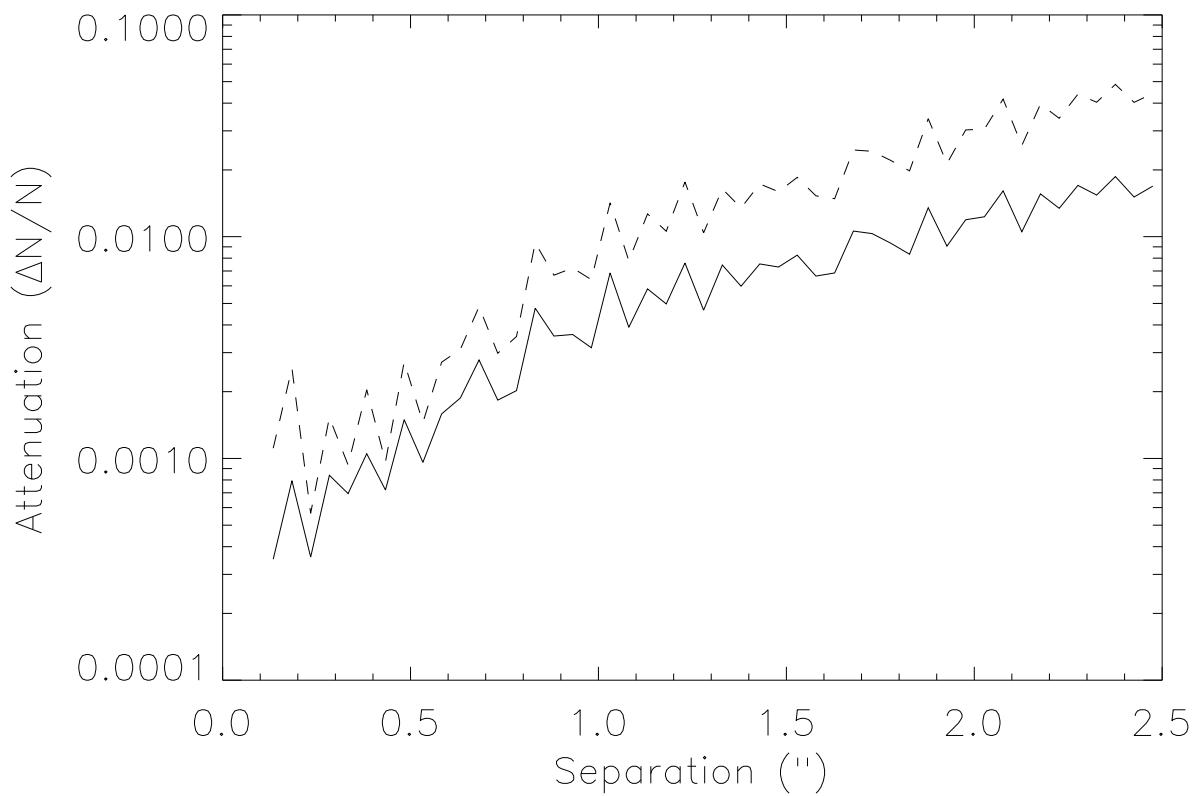}
\caption{PSF noise attenuation around an $m_H$ = 5 star observed for 10$^{5}$ second with an 8-m telescope, a high order AO system delivering a Strehl ratio of 0.91, a three-wavelength MCDA and 20\% global optical transmission without (solid line) and with (dashed line) a Lyot coronagraph. The latter features a gaussian mask with a FWHM of 15~$\lambda/D$ and a 60\% transmissive Lyot mask.\label{fig4}}
\end{figure}

\begin{figure}
\epsscale{1}
\plotone{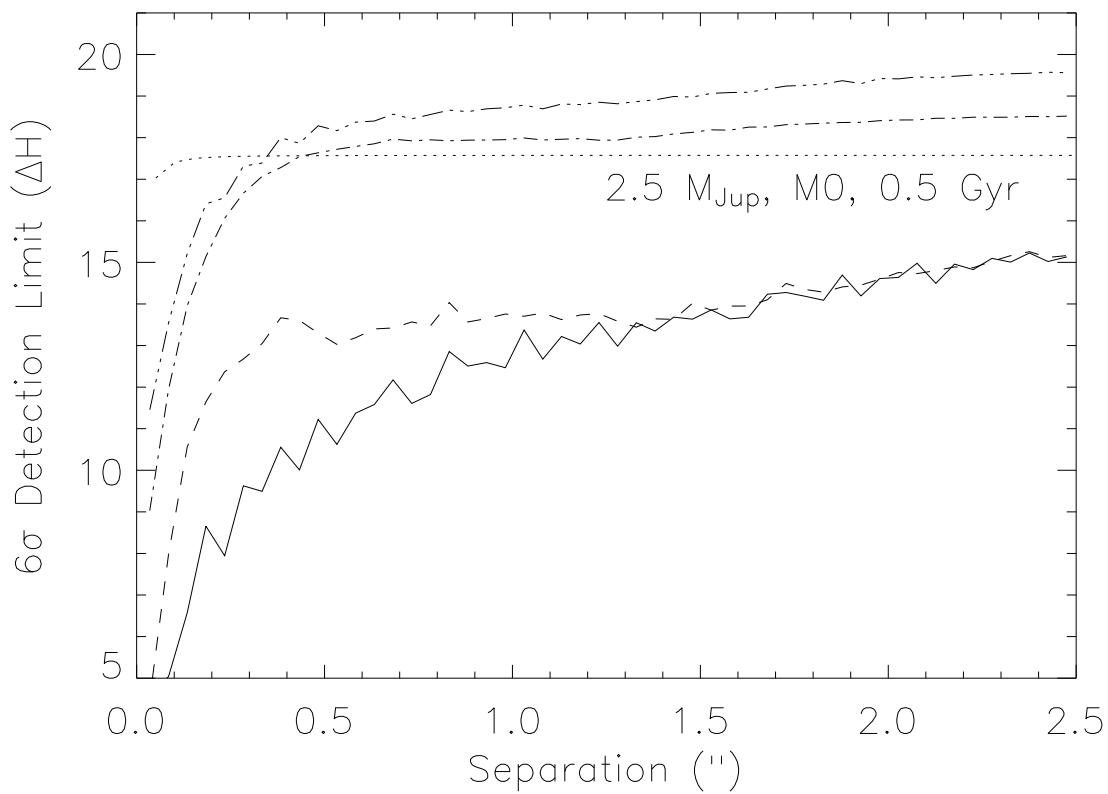}
\caption{$6 \sigma$ detection limit with (three dots dashed line) and without (dot dashed line) a Lyot coronagraph after multicolor speckle suppression. The dashed ans solid lines show respectively the noise of a single PSF with and without a Lyot coronagraph. The dotted line shows the magnitude difference for a 0.5~Gyr~/~2.5~M$_{\rm{Jup}}$ (from \citet{baraffe2003}) companion around a M0 star at 10~pc, including the effect of stellar flux reflection by the companion at small separations.\label{fig5}}
\end{figure}

Numerical simulations were done to estimate the PSF noise attenuation achievable with a three-wavelength ($\lambda_1 = 1.52$~$\mu$m, $\lambda_2 = 1.58$~$\mu$m and $\lambda_3 = 1.64$~$\mu$m) MCDA in quasi-realistic conditions. Simulations are for a 10$^5$~s total integration divided in 60~s exposures on an $m_H$~=~5 star observed with an 8-m telescope and 4k actuator AO system of 20\% transmission and 2\% bandpass. AO phase filtering is simulated by choosing a flat power spectrum for spatial frequencies lower than twice the interactuator spacing. Static aberrations are included using a power-law power spectrum with $\alpha = -2.7$ to obtain 100~nm RMS total aberrations, of which 15~nm RMS are not corrected by the AO system due to non-common path optics. The three images are the average of 500 atmospheric speckle realizations for 0.5$^{\prime \prime}$~seeing at 0.5~$\mu $m. AO filtering of both the static and Kolmogorov atmospheric phase screens yields PSFs with Strehl ratios of 0.96 at 1.60~$\mu $m. These also include a typical level of read noise of 5 e- per pixel per 60~s exposure, photon noise, random flat field errors of 10$^{-4}$, sky noise and sample to sample crosstalk of 10$^{-3}$. PSF evolution with wavelength is included and partly compensated by appropriate combination of the multi-wavelength images \citep{marois2000a,marois2004b}. For the sake of completeness, simulations were done both with and without an optimized Lyot coronagraph. For the coronagraph simulation, the 15~nm RMS unfiltered static aberration is split in two uncorrelated 10~nm RMS aberrations that are respectively included before and after the coronagraph. The results are shown in Fig.~\ref{fig4}. These lead to the detection limits (Fig.~\ref{fig5}), checked with off-axis companion images, which are compared to the contrast required for the detection of 2.5~M$_{\rm{Jup}}$ exoplanets around 0.5~Gyr M0 stars \citep{baraffe2003}. When combined with a MCDA, the coronagraph does $\sim 1$ magnitude better. The coronagraph has effectively attenuated the PSF structure by an order of magnitude inside a 0.5$^{\prime \prime}$~radius, thus reducing the flat field accuracy needed to achieve $\Delta H$~=~17.5.

\section{Summary and Conclusion}
Experience with the TRIDENT camera has shown that small non-common path aberrations in multiple optical channel instruments seriously limit PSF noise attenuation hence faint companion detection. The MCDA concept described here can overcome this problem. Numerical simulations predict that, with an MCDA, 2.5~M$_{\rm{Jup}}$ exoplanets around 0.5 Gyr M0 stars should be detectable in 10$^{5}$ seconds with an 8-m telescope and a 4k actuator AO system. 

Planet finders based on MCDAs would enable detection and characterization of exoplanets with present large telescopes and high order AO systems. The technique is relatively simple to implement and would complement coronagraph projects, or telescopes like HST, where it could calibrate evolving noise structure. An MCDA combined with a coronagraph would be a powerful tool for detecting ozone in the mid-IR \citep{desmarais2002}, water in the near-IR, molecular oxygen in the red and, eventually, the chlorophyll ``red edge'' with the Terrestrial Planet Finder. An optimized MCDA-type polarimeter could study reflected star light from exoplanets \citep{saar2003}, or disks and zodiacal dust. Instead of micro-filters, micro-polarizers would be used to produce simultaneous polarized images with a single optical channel.

\acknowledgments
This work was supported in part through grants from NSERC, Canada and from FQRNT, Qu\'{e}bec.

\clearpage
\clearpage 

\clearpage
\clearpage 

\clearpage
\clearpage 

\clearpage
\clearpage

\clearpage
\clearpage

\end{document}